\documentclass[showpacs,showkeys,amsmath,amssymb,preprint,11pt]{revtex4}
\usepackage{epsfig,dcolumn,bm}

\begin{document}

\title{Study on $N \bar{\Omega}$ systems in a chiral
quark model}

\author{D. Zhang$^{1,2}$}\thanks{E-mail: zhangdan@mail.ihep.ac.cn}
\author{F. Huang$^{3}$}
\author{L.R. Dai$^{4}$}
\author{Y.W. Yu$^1$}
\author{Z.Y. Zhang$^1$}\thanks{E-mail: zhangzy@mail.ihep.ac.cn}
\affiliation{\small
$^1$Institute of High Energy Physics, P.O. Box 918-4, Beijing 100049, PR China\footnote{Mailing address.} \\
$^2$Graduate School of the Chinese Academy of Sciences, Beijing, PR
China\\
$^3$CCAST (World Laboratory), P.O. Box 8730, Beijing 100080, PR
China\\
$^4$Department of Physics, Liaoning Normal University, Dalian
116029, PR China }


\begin{abstract}
The $N \bar{\Omega}$ systems with spin $S=1$ and $S=2$ are dynamically investigated within the framework
of the chiral SU(3) quark model and the extended chiral SU(3) quark model by solving the resonating group
method (RGM) equation. The model parameters are taken from our previous work, which gave a good
description of the energies of the baryon ground states, the binding energy of deuteron, and the
experimental data of the nucleon-nucleon ($NN$) and nucleon-hyperon ($NY$) scattering processes. The
results show that $N \bar{\Omega}$ states with spin $S=1$ and $S=2$ can be bound both in the chiral SU(3)
and extended chiral SU(3) quark models, and the binding energies are about $28-59$ MeV. When the
annihilation effect is considered, the binding energies increase to about $37-130$ MeV, which indicates
the annihilation effect plays a relatively important role in forming an $N \bar{\Omega}$ bound state. At
the same time, the $N \bar{\Omega}$ elastic scattering processes are also studied. The $S$, $P$, $D$
partial wave phase shifts and the total cross sections of $S=1$ and $S=2$ channels have been calculated
by solving the RGM equation for scattering problems.
\end{abstract}




\maketitle

\section{Introduction}

Baryon-antibaryon ($B\bar{B}$) system is believed to be a good field
to explore the quality of strong interactions, especially
short-range ones. Whether $N \bar{N}$ bound states or resonances
exist has been widely studied by a great deal of theoretical and
experimental scientists for a few decades, but up to now it is still
an open question. The main reason is that the annihilation effect at
short distance is very important in the $N\bar{N}$ system, which
enhances its complexity. From 1980s' processes of $N\bar{N}$
annihilation into two and three mesons were investigated  on quark
level and obtained some interesting results \cite{gree84, maru85,
maru87, dove92}. In their models, there were three kinds of
annihilation mechanisms: 1) the quark-antiquark ($q\bar{q}$) pair
could be destroyed and created with vacuum quantum numbers; 2)
quarks in $N$ and antiquarks in $\bar{N}$ rearrangement led to the
annihilation into mesons; 3) the $q\bar{q}$ pair annihilated with
the quantum number of gluon. And their analysis indicated that the
first one is the dominant among the three annihilation mechanisms in
the $N\bar{N}$ system, which can give a reasonable description of
$N\bar{N}$ annihilation data \cite{maru87, dove92}.

Despite that some progresses have been made in the study of the annihilation effect, there are still some
uncertainties in the $N\bar{N}$ systems, because in which there are three different annihilation modes
and it is difficult to distinguish the contribution and characteristic of each mechanism. Thus it is hard
to give a convinced theoretical prediction for $N\bar{N}$ bound states or resonances. It seems more
appropriate to choose some special systems which have only one kind of annihilation mechanism. We think
the $N\bar{\Omega}$ system is an interesting one. Since $\bar{\Omega}$ is composed of three $\bar{s}$
quarks and $N$ of three $u(d)$ quarks, $u\bar{s}$ or ${d\bar{s}}$ is impossible to annihilate to the
vacuum, and also impossible to annihilate to gluon because gluon is flavorless. Therefore, the
$N\bar{\Omega}$ system can only annihilate into three mesons with strangeness by rearrangement, and  it
provides an optimal place to study the rearrangement mechanism of the annihilation processes. Moreover,
for the $t$-channel interactions, there is no one gluon exchange interaction and the meson exchanges play
important roles in this system, so it is also an ideal place for examining the chiral field coupling.

As is well known, the quantum chromodynamics (QCD) is the underlying theory of the strong interaction. At
high-energy region, the perturbative treatment of QCD is quit successful, while it fails at low and
intermedium energy domain. However, nonperturbative QCD (NPQCD) effect is very important for light quark
systems and till now there is no serious approach to solve the NPQCD problem. To study the baryon
physics, people still need QCD-inspired models to help. Among these models, the chiral SU(3) quark model
has been quite successful in reproducing the energies of the baryon ground states, the binding energy of
the deuteron, the nucleon-nucleon ($NN$) scattering phase shifts, and the hyperon-nucleon ($YN$) cross
sections \cite{zyzhang97}. In this model, the quark-quark interaction containing confinement, one gluon
exchange (OGE) and boson exchanges stemming from chiral-quark coupling. In the study of $NN$ interactions
on quark level, the short-range feature can be explained by OGE interaction and the quark exchange
effect. As we know, in the traditional one boson exchange (OBE) model on baryon level, the short-range
$NN$ interaction comes from vector meson ($\rho,\omega, K^*$ and $\phi$) exchanges. In order to study
vector-meson exchange effect on quark level, the extended chiral SU(3) quark model was proposed
\cite{lrdai03}. In this extended model, we further include the coupling of the quark and vector chiral
fields. The OGE that acts an important role in the short-range quark-quark interaction in the chiral
SU(3) quark model is now nearly replaced by the vector meson exchanges in the extended chiral SU(3) quark
model. This extended chiral quark model can also reasonably explain the energies of the baryon ground
states, the binding energy of the deuteron, and the $NN$ scattering phase shifts \cite{lrdai03}. Recently
both the chiral SU(3) quark model and the extended chiral SU(3) quark model have been extended to the
systems with antiquarks to study the baryon-meson interactions by solving a resonating group method (RGM)
equation \cite{fhuang}. Some interesting results were obtained, which are quite similar to those given by
the chiral unitary approach study \cite{nkai}. Inspired by all these achievements, we try to extend our
study to the baryon-antibaryon systems in the framework of these two models.

In the present work, we dynamically investigate the characteristic
of the $N\bar{\Omega}$ systems with spin $S=1$ and $S=2$ in the
chiral SU(3) quark model and the extended SU(3) quark model. The
model parameters are taken from our previous work \cite{zyzhang97,
lrdai03}. Firstly, the binding energies of the $N\bar{\Omega}$
states are studied, and the annihilation effect is discussed as
well. The results show that $N \bar{\Omega}$ states with spin
$S=1$ and $S=2$ can be bound states both in the chiral SU(3) and
extended chiral SU(3) quark models, and the binding energies range
from $28$ MeV to $59$ MeV. When the annihilation effect is
considered, the binding energies increase to around $37-130$ MeV.
Secondly,  in order to get more information of the $N\bar{\Omega}$
structure,  the $N \bar{\Omega}$ elastic scattering processes are
also studied and the phase shifts of $S$, $P$ and $D$ partial
waves and the total cross sections are obtained.

The paper is organized as follows. In the next section the framework
of the chiral SU(3) quark model and the extended chiral SU(3) quark
model are briefly introduced. The calculated results of the
$N\bar{\Omega}$ states are shown in Sec. III, where some discussions
are made as well. Finally, the summary is given in Sec. IV.

\section{Formulation}

\subsection{Model}

The  chiral SU(3) quark model and the extended chiral SU(3) quark
model have been widely described in the literature
\cite{zyzhang97,lrdai03} and we refer the reader to those works for
details. Here we just give the salient feature of these two models.

In these two models, the total Hamiltonian of baryon-antibaryon
systems can be written as
\begin{eqnarray}
\label{hami6q} H=\sum_{i=1}^6 T_{i}-T_{G}+\sum_{i<j=1}^3 V_{qq}(ij)+
\sum_{i<j=4}^6 V_{\bar q \bar{q}}(ij)+
\sum_{\genfrac{}{}{0pt}{}{i=1,3}{j=4,6}} V_{q \bar q} (ij),
\end{eqnarray}
where $T_G$ is the kinetic energy operator for the center-of-mass
motion, and $V_{qq}(ij)$ represents the interaction between two
quarks ($qq$),

\begin{eqnarray}
V_{qq}(ij)= V^{OGE}_{qq}(ij) + V^{conf}_{qq}(ij) + V^{ch}_{qq}(ij),
\end{eqnarray}
where $V^{OGE}_{qq}(ij)$ is the one-gluon-exchange interaction,
\begin{eqnarray}
\label{oge}
V^{OGE}_{qq}(ij)&=&\frac{1}{4}g_{i}g_{j}\left(\lambda^c_i\cdot\lambda^c_j\right)\nonumber\\
&&\times\left\{\frac{1}{r_{ij}}-\frac{\pi}{2} \delta({\bm r}_{ij})
\left(\frac{1}{m^2_{q_i}}+\frac{1}{m^2_{q_j}}+\frac{4}{3}\frac{1}{m_{q_i}m_{q_j}}
({\bm \sigma}_i \cdot {\bm \sigma}_j)\right)\right\}+V_{OGE}^{\bm l
\cdot \bm s}++V_{OGE}^{ten},
\end{eqnarray}
and $V^{conf}_{qq}(ij)$ is the confinement potential, taken as the
quadratic form,
\begin{eqnarray}
\label{conf}
V^{conf}_{qq}(ij)=-a_{ij}^{c}(\lambda_{i}^{c}\cdot\lambda_{j}^{c})r_{ij}^2
-a_{ij}^{c0}(\lambda_{i}^{c}\cdot\lambda_{j}^{c}).
\end{eqnarray}
$V^{ch}_{qq}(ij)$ represents the chiral fields induced effective
quark-quark potential. In the chiral SU(3) quark model,
$V^{ch}_{ij}$ includes the scalar boson exchanges and the
pseudoscalar boson exchanges,
\begin{eqnarray}
V^{ch}_{qq}(ij) = \sum_{a=0}^8 V_{\sigma_a}({\bm
r}_{ij})+\sum_{a=0}^8 V_{\pi_a}({\bm r}_{ij}),
\end{eqnarray}
and in the extended chiral SU(3) quark model, the vector boson
exchanges are also included,
\begin{eqnarray}
V^{ch}_{qq}(ij) = \sum_{a=0}^8 V_{\sigma_a}({\bm
r}_{ij})+\sum_{a=0}^8 V_{\pi_a}({\bm r}_{ij})+\sum_{a=0}^8
V_{\rho_a}({\bm r}_{ij}).
\end{eqnarray}
Here $\sigma_{0},...,\sigma_{8}$ are the scalar nonet fields,
$\pi_{0},..,\pi_{8}$ the pseudoscalar nonet fields, and
$\rho_{0},..,\rho_{8}$ the vector nonet fields. The expressions of
these potentials are
\begin{eqnarray}
V_{\sigma_a}({\bm r}_{ij})=-C(g_{ch},m_{\sigma_a},\Lambda)
X_1(m_{\sigma_a},\Lambda,r_{ij}) [\lambda_a(i)\lambda_a(j)] +
V_{\sigma_a}^{\bm {l \cdot s}}({\bm r}_{ij}),
\end{eqnarray}
\begin{eqnarray}
V_{\pi_a}({\bm r}_{ij})=C(g_{ch},m_{\pi_a},\Lambda)
\frac{m^2_{\pi_a}}{12m_{q_i}m_{q_j}} X_2(m_{\pi_a},\Lambda,r_{ij})
({\bm \sigma}_i\cdot{\bm \sigma}_j) [\lambda_a(i)\lambda_a(j)]
+V_{\pi_a}^{ten}({\bm r}_{ij}),
\end{eqnarray}
\begin{eqnarray}
V_{\rho_a}({\bm r}_{ij})&=&C(g_{chv},m_{\rho_a},\Lambda)\left\{
X_1(m_{\rho_a},\Lambda,r_{ij})+ \frac{m^2_{\rho_a}}{6m_{q_i}m_{q_j}}
\left(1+\frac{f_{chv}}{g_{chv}}\frac{m_{q_i}+m_{q_j}}{M_P}+\frac{f^2_{chv}}{g^2_{chv}}
\right. \right. \nonumber \\
&& \left. \left. \times \frac{m_{q_i}m_{q_j}}{M^2_P}\right)
X_2(m_{\rho_a},\Lambda,r_{ij})({\bm \sigma}_i\cdot{\bm \sigma}_j)
\right\}[\lambda_a(i)\lambda_a(j)] + V_{\rho_a}^{\bm {l \cdot
s}}({\bm r}_{ij}) + V_{\rho_a}^{ten}({\bm r}_{ij}),
\end{eqnarray}
where
\begin{eqnarray}
C(g_{ch},m,\Lambda)=\frac{g^2_{ch}}{4\pi}
\frac{\Lambda^2}{\Lambda^2-m^2} m,
\end{eqnarray}
\begin{eqnarray}
\label{x1mlr} X_1(m,\Lambda,r)=Y(mr)-\frac{\Lambda}{m} Y(\Lambda r),
\end{eqnarray}
\begin{eqnarray}
X_2(m,\Lambda,r)=Y(mr)-\left(\frac{\Lambda}{m}\right)^3 Y(\Lambda
r),
\end{eqnarray}
\begin{eqnarray}
Y(x)=\frac{1}{x}e^{-x},
\end{eqnarray}
and $M_P$ being a mass scale, taken as proton mass. $m_{\sigma_a}$
is the mass of the scalar meson, $m_{\pi_a}$ the mass of the
pseudoscalar meson, and $m_{\rho_a}$ the mass of the vector meson.

$V_{\bar q\bar q}(ij)$ in Eq.(\ref{hami6q}) is the
antiquark-antiquark ($\bar q \bar q$) interaction,
\begin{eqnarray}
V_{\bar q \bar q}=V_{\bar q\bar q}^{conf}+V_{\bar q\bar
q}^{OGE}+V_{\bar q\bar q}^{ch}.
\end{eqnarray}
Replacing the $\lambda^c_i\cdot\lambda^c_j$ in Eq.(\ref{oge}) and
Eq.(\ref{conf}) by $\lambda^{c\ast}_i\cdot\lambda^{c\ast}_j$, we can
obtain the forms of $V_{\bar q\bar q}^{OGE}$ and $V_{\bar q\bar
q}^{conf}$. $V_{\bar q\bar q}^{ch}$ has the same form as
$V_{qq}^{ch}$.

$V_{q \bar q}(ij)$ in Eq.(\ref{hami6q}) represents the interaction
between quark and antiquark ($q\bar{q}$). Between $N$ and
$\bar{\Omega}$,
there is no one-gluon-exchange interaction and the confinement
potential scarcely contributes any interactions, so $V_{q \bar q}$
only includes two parts: the meson fields induced effective
potential and annihilation parts,
\begin{equation}
V_{q \bar{q}}=V_{q\bar q}^{ch}+V^{ann}_{q\bar q},
\end{equation}
where
\begin{equation}
V_{q\bar{q}}^{ch}=\sum_{j}(-1)^{G_j}V_{qq}^{ch,j}.
\end{equation}
Here $(-1)^{G_j}$ represents the G parity of the $j$th meson. As
mentioned above, for the $N \bar{\Omega}$ systems $u(d)\bar{s}$ can
only annihilate into $K$ and $K^*$ mesons by rearrangement
mechanism---i.e.,
\begin{eqnarray}
V_{q\bar q}^{ann}=V_{ann}^{K}+V_{ann}^{K^*},
\end{eqnarray}
with
\begin{eqnarray}
V_{ann}^{K}=C^K\left(\frac{1-{\bm \sigma}_q \cdot {\bm
\sigma}_{\bar{q}}}{2}\right)_{s}\left(\frac{2 + 3\lambda_q \cdot
\lambda^*_{\bar{q}}}{6}\right)_{c} \left(\frac{38+3\lambda_q \cdot
\lambda^*_{\bar q}}{18}\right)_{f}\delta({\bm r}),
\end{eqnarray}
and
\begin{eqnarray}
V_{ann}^{K^*}=C^{K^*}\left(\frac{3+{\bm \sigma}_q \cdot {\bm
\sigma}_{\bar{q}}}{2}\right)_{s}\left(\frac{2 + 3\lambda_q \cdot
\lambda^*_{\bar{q}}}{6}\right)_{c} \left(\frac{38+3\lambda_q \cdot
\lambda^*_{\bar q}}{18}\right)_{f}\delta({\bm r}),
\end{eqnarray}
where $C^K$ and $C^{K^*}$ are treated as parameters and we adjust
them to fit the masses of $K$ and $K^*$ mesons.

\subsection{Determination of the parameters}

{\small
\begin{table}[htb]
\caption{\label{para} Model parameters. The meson masses and the
cutoff masses: $m_{\sigma'}=980$ MeV, $m_{\kappa}=980$ MeV,
$m_{\epsilon}=980$ MeV, $m_{\pi}=138$ MeV, $m_K=495$ MeV,
$m_{\eta}=549$ MeV, $m_{\eta'}=957$ MeV, $m_{\rho}=770$ MeV,
$m_{K^*}=892$ MeV, $m_{\omega}=782$ MeV, $m_{\phi}=1020$ MeV, and
$\Lambda=1100$ MeV for all mesons.}
\begin{center}
\begin{tabular}{cccc}
\hline\hline
  & Chiral SU(3) quark model & \multicolumn{2}{c}{Extended chiral SU(3) quark model}  \\
  &   I   &    II    &    III \\  \cline{3-4}
  &  & $f_{chv}/g_{chv}=0$ & $f_{chv}/g_{chv}=2/3$ \\
\hline
 $b_u$ (fm)  & 0.5 & 0.45 & 0.45 \\
 $m_u$ (MeV) & 313 & 313 & 313 \\
 $m_s$ (MeV) & 470 & 470 & 470 \\
 $g_u^2$     & 0.766 & 0.056 & 0.132 \\
 $g_s^2$     & 0.846 & 0.203 & 0.250 \\
 $g_{ch}$    & 2.621 & 2.621 & 2.621  \\
 $g_{chv}$   &       & 2.351 & 1.973  \\
 $m_\sigma$ (MeV) & 595 & 535 & 547 \\
 $a^c_{uu}$ (MeV/fm$^2$) & 46.6 & 44.5 & 39.1 \\
 $a^c_{us}$ (MeV/fm$^2$) & 58.7 & 79.6 & 69.2 \\
 $a^c_{ss}$ (MeV/fm$^2$) & 99.2 & 163.7 & 142.5 \\
 $a^{c0}_{uu}$ (MeV)  & $-$42.4 & $-$72.3 & $-$62.9 \\
 $a^{c0}_{us}$ (MeV)  & $-$36.2 & $-$87.6 & $-$74.6 \\
 $a^{c0}_{ss}$ (MeV)  & $-$33.8 & $-$108.0 & $-$91.0 \\
\hline\hline
\end{tabular}
\end{center}
\end{table}}

All the model parameters are taken from our previous work
\cite{zyzhang97,lrdai03}, which can give a satisfactory description
of the energies of the baryon ground states, the binding energy of
deuteron, the $NN$ scattering phase shifts. The harmonic-oscillator
width parameter $b_u$ is taken with different values for the two
models: $b_u=0.50$ fm in the chiral SU(3) quark model and $b_u=0.45$
fm in the extended chiral SU(3) quark model. This means that the
bare radius of baryon becomes smaller when more meson clouds are
included in the model, which sounds reasonable in the sense of the
physical picture. The up (down) quark mass $m_{u(d)}$ and the
strange quark mass $m_s$ are taken to be the usual values:
$m_{u(d)}=313$ MeV and $m_s=470$ MeV. The coupling constant for
scalar and pseudoscalar chiral field coupling, $g_{ch}$, is
determined according to the relation
\begin{eqnarray}
\frac{g^{2}_{ch}}{4\pi} = \left( \frac{3}{5} \right)^{2}
\frac{g^{2}_{NN\pi}}{4\pi} \frac{m^{2}_{u}}{M^{2}_{N}},
\end{eqnarray}
with empirical value $g^{2}_{NN\pi}/4\pi=13.67$. The coupling
constant for vector coupling of the vector-meson field is taken to
be $g_{chv}=2.351$, the same as used in the $NN$ case
\cite{lrdai03}. The masses of the mesons are taken to be the
experimental values, except for the $\sigma$ meson. The $m_\sigma$
is adjusted to fit the binding energy of the deuteron. The cutoff
radius $\Lambda^{-1}$ is taken to be the value close to the chiral
symmetry breaking scale \cite{ito90}. The OGE coupling constants,
$g_{u}$ and $g_{s}$, can be determined by the mass splits between
$N$, $\Delta$ and $\Sigma$, $\Lambda$, respectively. The confinement
strengths $a^{c}_{uu}$, $a^{c}_{us}$, and $a^{c}_{ss}$ are fixed by
the stability conditions of $N$, $\Lambda$, and $\Xi$ and the
zero-point energies $a^{c0}_{uu}$, $a^{c0}_{us}$, and $a^{c0}_{ss}$
by fitting the masses of $N$, $\Sigma$, and $\overline{\Xi+\Omega}$,
respectively. All the parameters are tabulated in Table \ref{para},
where the first set is for the chiral SU(3) quark model (I), the
second and third sets are for the extended chiral SU(3) quark model
by taking $f_{chv}/g_{chv}$ as $0$ (II) and $2/3$ (III),
respectively. Here $g_{chv}$ and $f_{chv}$ are the coupling
constants for vector coupling and tensor coupling of the vector
meson fields, respectively.

\subsection{Resonating group method (RGM)}

With all the parameters determined, the $N \bar{\Omega}$ system can
be dynamically studied in the framework of the RGM, a well
established method for detecting the interaction between two
clusters. The cases for the $N\bar{\Omega}$ states are much simpler,
since there are three quarks in $N$ and three antiquarks in
$\bar{\Omega}$, and antisymmetrization between $N$ and
$\bar{\Omega}$ is not necessary. Thus the wave function of this
six-quark system is taken as
\begin{eqnarray}
\Psi={\hat \phi}_N(\bm \xi_1,\bm \xi_2){\hat \phi}_{\bar{\Omega}}(\bm \xi_3,\bm \xi_4)\chi({\bm
R}_{N\bar{\Omega}}),
\end{eqnarray}
where ${\bm \xi}_1$, ${\bm \xi}_2$ are the internal coordinates for
the cluster $N$, and ${\bm \xi}_3$, ${\bm \xi}_4$ the internal
coordinate for the cluster $\bar{\Omega}$. ${\bm R}_{N
\bar{\Omega}}\equiv {\bm R}_N-{\bm R}_{\bar{\Omega}}$ is the
relative coordinate between the two clusters, $N$ and
$\bar{\Omega}$. The ${\hat \phi}_N$ $({\hat \phi}_{\bar{\Omega}})$
is the antisymmetrized internal cluster wave function of $N$
$(\bar{\Omega})$, and $\chi ({\bm R}_{N \bar{\Omega}})$ the relative
wave function of the two clusters. For a bound-state problem or a
scattering one, by solving the equation
\begin{equation}
\langle \delta\Psi|(H-E)|\Psi \rangle=0,
\end{equation}
we can  obtain binding energies or scattering phase shifts for the
two-cluster systems. The details of solving the RGM equation can be
found in Refs. \cite{kwi77,mka77,fhuang}.

\section{Results and discussions}

\begin{figure}[htb]
\epsfig{file=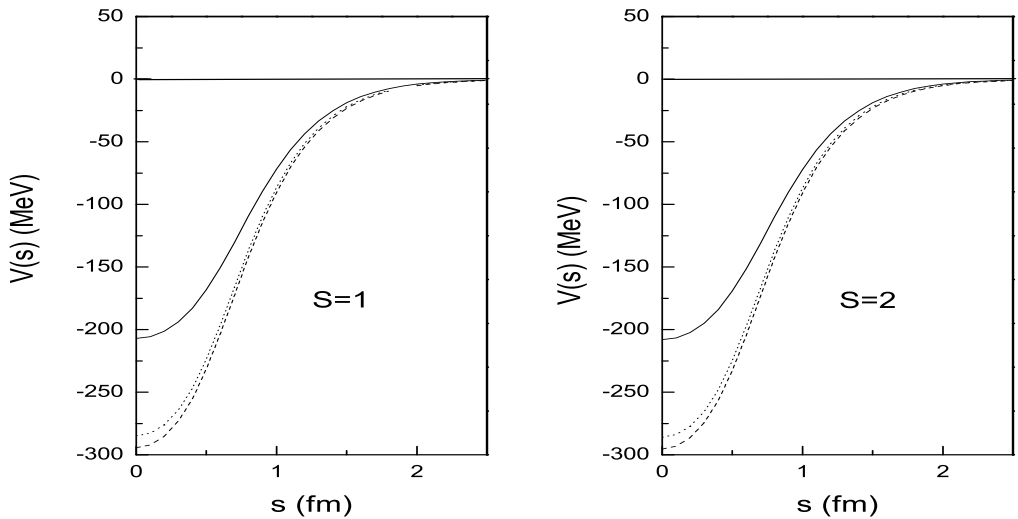,width=15.0cm,height=8.0cm} \vglue -0.5cm \caption{\small \label{poten} The GCM matrix
elements of the $S$-wave $N\bar{\Omega}$ effective potential as a function of the generator coordinate,
where the annihilation effect is not included. The solid line represents the results obtained in chiral
SU(3) quark model with set I, and the dashed and dotted lines represent the results in extended chiral
SU(3) quark model with set II and set III, respectively. }
\end{figure}

{\small
\begin{table}[htb]
\caption{{\label{bind}}Binding energy $B_{N \bar{\Omega}}$ and
corresponding RMS radius $\overline{R}$ of $N \bar{\Omega}$ without
the annihilation effect.}
\begin {center}
\begin{tabular*}{140mm}{@{\extracolsep\fill}lcccc}
\hline\hline Model & \multicolumn{2}{c}{S=1}&
\multicolumn{2}{c}{S=2}\\  \cline{2-3}\cline{4-5}
 &$B_{N \bar{\Omega}}$ (MeV)&$\overline{R}$ (fm) &$B_{N \bar{\Omega}}$ (MeV)&$\overline{R}$ (fm) \\
\hline
  I   &28.3& 0.8&28.8& 0.8\\
II &58.8& 0.7&59.5& 0.7\\
III & 53.7& 0.7&54.4& 0.7\\
\hline\hline
\end{tabular*}
\end{center}
\end{table}}

The bound-state case of $S$-wave $N \bar{\Omega}$ systems with spin $S=1$ and $S=2$ is investigated in
both the chiral SU(3) quark model and the extended chiral SU(3) quark model.  As the first step, we will
not consider the annihilation effect. Fig. \ref{poten} shows the diagonal matrix elements of the
interaction potentials for the $N\bar{\Omega}$ systems with $S=1$ and $S=2$ in the generator coordinate
method (GCM) \cite{kwi77} calculations, which can be regarded as the effective potential of two clusters
$N$ and $\bar{\Omega}$. In Fig. \ref{poten}, $V(s)$ denotes the effective potential between $N$ and
$\bar{\Omega}$, and $s$ denotes the generator coordinate which can qualitatively describe the distance
between the two clusters. From Fig. \ref{poten}, we can see that for both $S=1$ and $S=2$ states,
effective potentials are attractive, and the attractions in the extended chiral SU(3) quark model are
greater than those in the chiral SU(3) quark model. Since the $N \bar{\Omega}$ system is quite special,
in both the chiral SU(3) quark model and the extended chiral SU(3) quark model, there is no OGE and no
$\sigma'$, $\kappa$, $\pi$, $K$, $\rho$, $K^*$, $\omega$, $\phi$ exchanges between $N$ and
$\bar{\Omega}$, thus the attractive force between them is mainly from $\sigma$ exchange. In our
calculation, the model parameters are fitted by the $NN$ scattering phase shifts and the mass of $\sigma$
is adjusted by fitting the deuteron's binding energy, so the value of $m_{\sigma}$ is somewhat different
for these three cases. In sets II and III, the mass of $\sigma$ meson is smaller than that of sets I, so
the $N\bar{\Omega}$ states can get more attractions in the extended chiral SU(3) quark model. Meanwhile,
the results of $N\bar{\Omega}$ with spin $S=1$ and $S=2$ are quite similar. This is easy to be
understood, because as presented above, in the $S$-wave calculation, the $\sigma$ exchange plays the
dominant role, which is spin independent.

By solving bound-state RGM equation, the calculated binding energies
and corresponding root-mean-squared (RMS) radii are obtained and
tabulated in Table \ref{bind}. One can see  that such attractive
potentials can make for bound states of the $N\bar{\Omega}$ systems.
Here, the binding energy ($B_{N \bar{\Omega}}$) and RMS radius
($\overline{R}$) are defined as
\begin{equation}
B_{N \bar{\Omega}}=-[ E_{N \bar{\Omega}}-(M_N+M_{\bar{\Omega}}) ]
\end{equation}
\begin{equation}
\overline{R}=\sqrt{\frac{1}{6} \sum_{i=1}^6 \langle (\bm {r_i}-\bm
{R_{cm}})^2 \rangle}
\end{equation}
From Table \ref{bind}, we see that, for both $S=1$ and $S=2$ channels, the binding energies of
$N\bar{\Omega}$ bound states are about 28 MeV in set I, i.e., in the chiral SU(3) quark model, and about
59 MeV in Set II, i.e., in the extended chiral SU(3) quark model with $f_{chv}/g_{chv}=0$, and about 54
MeV in set III, i.e., in the extended chiral SU(3) quark model with $f_{chv}/g_{chv}=2/3$. As have seen
in Fig. \ref{poten}, the $N\bar{\Omega}$ interactions for both $S=1$ and $S=2$ are more attractive in the
extended chiral SU(3) quark model than those in the chiral SU(3) quark model, and thus the sets II and
III get bigger binding energies than those of set I.

\begin{figure}[htb]
\epsfig{file=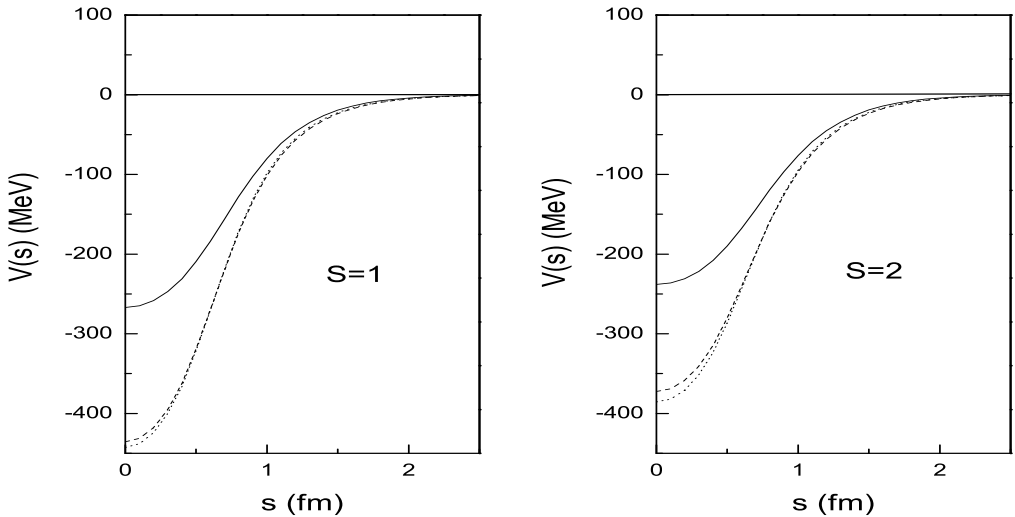,width=15.0cm,height=8.0cm} \vglue -0.5cm \caption{\small \label{potenann} The GCM
matrix elements of the $S$-wave $N\bar{\Omega}$ effective potential as a function of the generator
coordinate, where the annihilation effect is included. Same notation as in Fig. \ref{poten}.}
\end{figure}

{\small
\begin{table}[htb]
\caption{{\label{bindann}}Binding energy $B_{N \bar{\Omega}}$ and
corresponding RMS radius $\overline{R}$ of $N \bar{\Omega}$ with the
annihilation effect.}
\begin{center}
\begin{tabular*}{140mm}{@{\extracolsep\fill}lcccc}
\hline\hline Model & \multicolumn{2}{c}{S=1}&
\multicolumn{2}{c}{S=2}\\  \cline{2-3}\cline{4-5}
 &$B_{N \bar{\Omega}}$ (MeV)&$\overline{R}$ (fm) &$B_{N \bar{\Omega}}$ (MeV)&$\overline{R}$ (fm) \\
\hline
  I &46.2& 0.8&37.2& 0.8\\
II& 129.5& 0.6&102.0& 0.6\\
III & 132.2& 0.6&107.3& 0.6\\
\hline\hline
\end{tabular*}
\end{center}
\end{table}}

Compared with the results of $(N\Omega)_{S=2}$ system
\cite{lrdai06}, in which the predicted binding energies are 3.0
MeV in set I, 20.4 MeV in set II and 12.1 MeV in set III,
respectively, the binding energies of $(N\bar{\Omega})_{S=2}$ are
larger for all of these three cases. This is
because $K$ and $\kappa$ meson exchanges provide repulsive
interactions in the $(N\Omega)_{S=2}$ system, while they have no
contribution in the $(N\bar{\Omega})_{S=2}$ case, thus
$(N\bar{\Omega})_{S=2}$ can get more binding energies.

The root-mean-squared radius for the states of $(N\bar{\Omega})_{S=1}$ and  $(N\bar{\Omega})_{S=2}$ are
also calculated. In Table \ref{bind}, the RMS radii we acquired  are relatively small ($\sim$0.7-0.8 fm),
it seems that the annihilation effect should be considered in our calculation for these two states. As
pointed out above, for the $N\bar{\Omega}$ system, the annihilations to vacuum and gluon are forbidden
and the $u\bar{s}$($d\bar{s}$) can only annihilate to $K$ and $K^{\ast}$ mesons. Therefore, in the
following calculations, we will consider the annihilation to $K$ and $K^{\ast}$ mesons by using the Eqs.
(17)-(19), and the parameters $C^K$ and $C^{K^*}$ in Eqs. (18) and (19) are fitted by the masses of $K$
and $K^*$ mesons.

After including the annihilation effect, the diagonal GCM matrix elements of the effective potentials for
both $S=1$ and $S=2$ $N\bar{\Omega}$ states  are illustrated in Fig. \ref{potenann}, and the numerical
results of the binding energies and corresponding RMS radii with the annihilation effect involved are
shown in Table \ref{bindann}. Obviously for both $S=1$ and $S=2$, the effective potential become more
attractive and the binding energies increase after considering the annihilation effect. In set I, the
energy shift is about 18 MeV for $S=1$ and 8 MeV for $S=2$. In set II, it is about 70 MeV  for $S=1$ and
42 MeV for $S=2$, and in set III, about 78 MeV for $S=1$ and 54 MeV for $S=2$. It seems that, after the
annihilation effect is included in the $N\bar{\Omega}$ systems, the $N\bar{\Omega}$ systems can be
regarded as deeply bound states, especially in the extended chiral SU(3) quark model.

Additionally, the binding energies of $P$-wave $N \bar{\Omega}$ system are also studied. The results
manifest that no matter whether the annihilation effect is taken into account, $P$-wave $N \bar{\Omega}$
systems are always unbound in both the chiral SU(3) quark model and the extended chiral SU(3) quark
model.

\begin{figure}[htb]
\epsfig{file=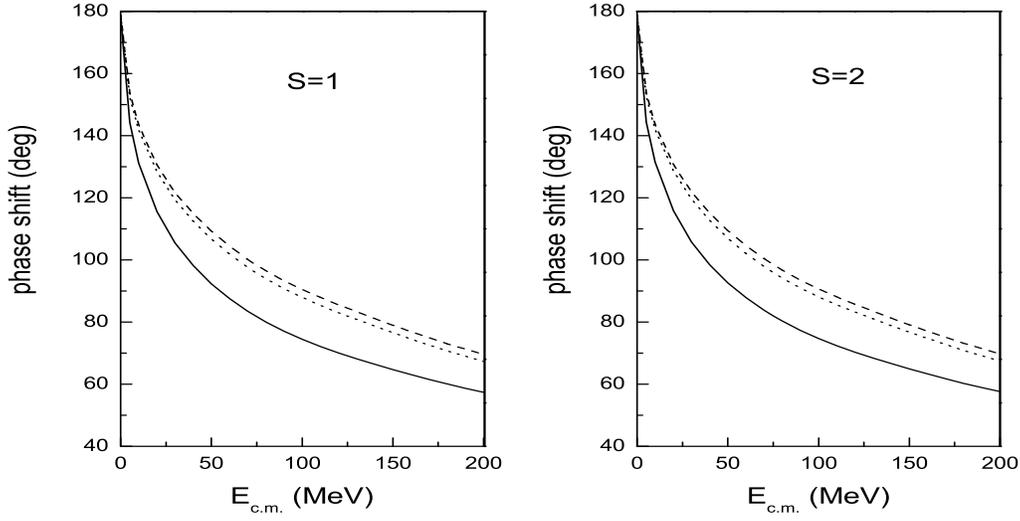,width=15.0cm,height=8.0cm} \vglue -0.5cm \caption{\small \label{phase1} $N
\bar{\Omega}$ $S$-wave phase shifts as a function of the energy of the center of mass motion. Same
notation as in Fig. \ref{poten}. }
\end{figure}

\begin{figure}[htb]
\epsfig{file=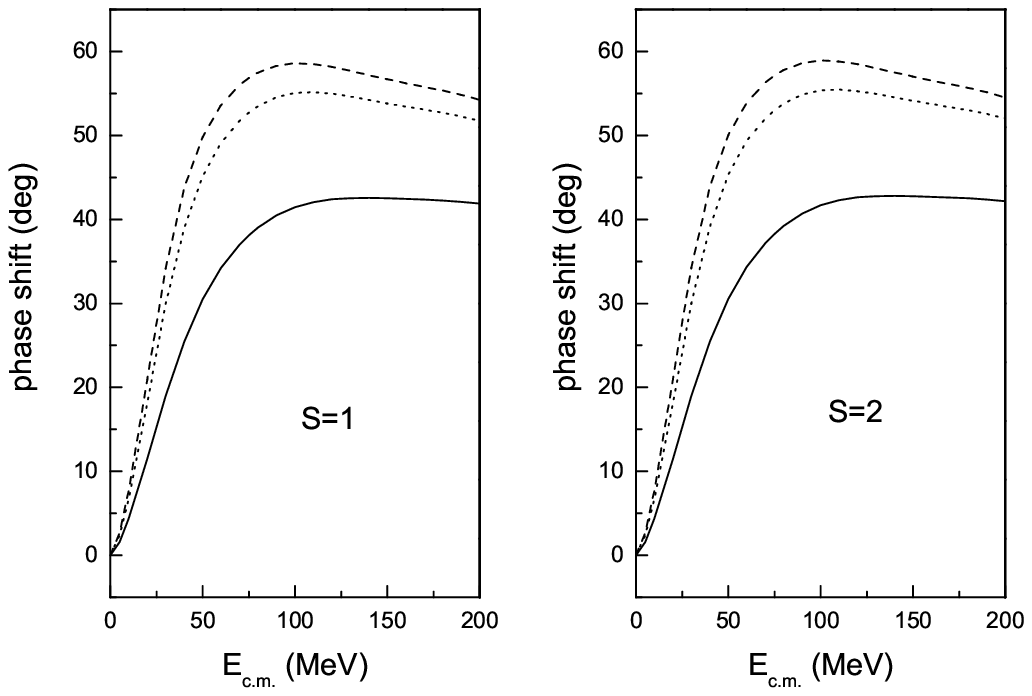,width=15.0cm,height=8.0cm} \vglue -0.5cm \caption{\small \label{phase2} $N
\bar{\Omega}$ $P$-wave phase shifts as a function of the energy of the center of mass motion. Same
notation as in Fig. \ref{poten}.}
\end{figure}

\begin{figure}[htb]
\epsfig{file=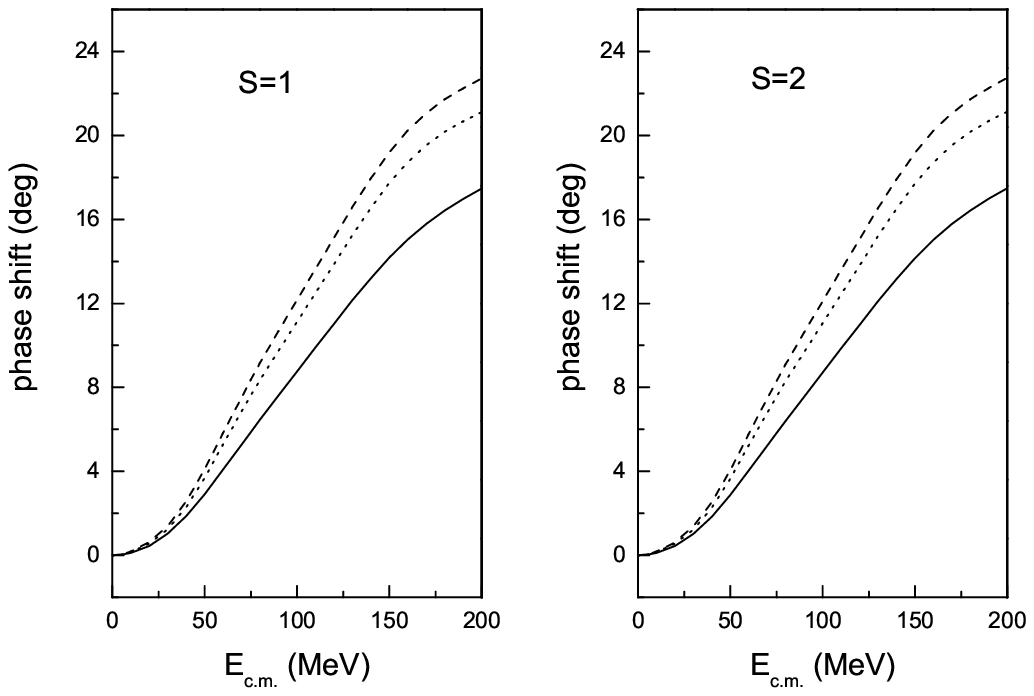,width=15.0cm,height=8.0cm} \vglue -0.5cm \caption{\small \label{phase3} $N
\bar{\Omega}$ $D$-wave phase shifts as a function of the energy of the center of mass motion. Same
notation as in Fig. \ref{poten}.}
\end{figure}

To get more information about the systems of $N \bar{\Omega}$, we further study the $N \bar{\Omega}$
elastic scattering processes. The phase shifts of $S$, $P$ and $D$ partial waves of
$(N\bar{\Omega})_{S=2}$ and $(N\bar{\Omega})_{S=1}$ are calculated. As a primary study, the spin-orbit
and tensor forces are not considered for the $P$ and $D$ waves, i.e. only central force is considered.
The phase shifts of $(N\bar{\Omega})_{S=2}$ and $(N\bar{\Omega})_{S=1}$ for the $S$, $P$ and $D$ partial
waves are illustrated in Fig. \ref{phase1}, Fig. \ref{phase2} and Fig. \ref{phase3}, respectively. We see
that the signs of the phase shifts in these two models are the same, and the magnitudes of the phase
shifts in the extended chiral SU(3) quark model are higher, especially for set II. This indicates that
the $N\bar{\Omega}$ systems get more attractive interactions in the extended chiral SU(3) quark model,
consisted with the results of the binding energy calculation.

\begin{figure}[htb]
\epsfig{file=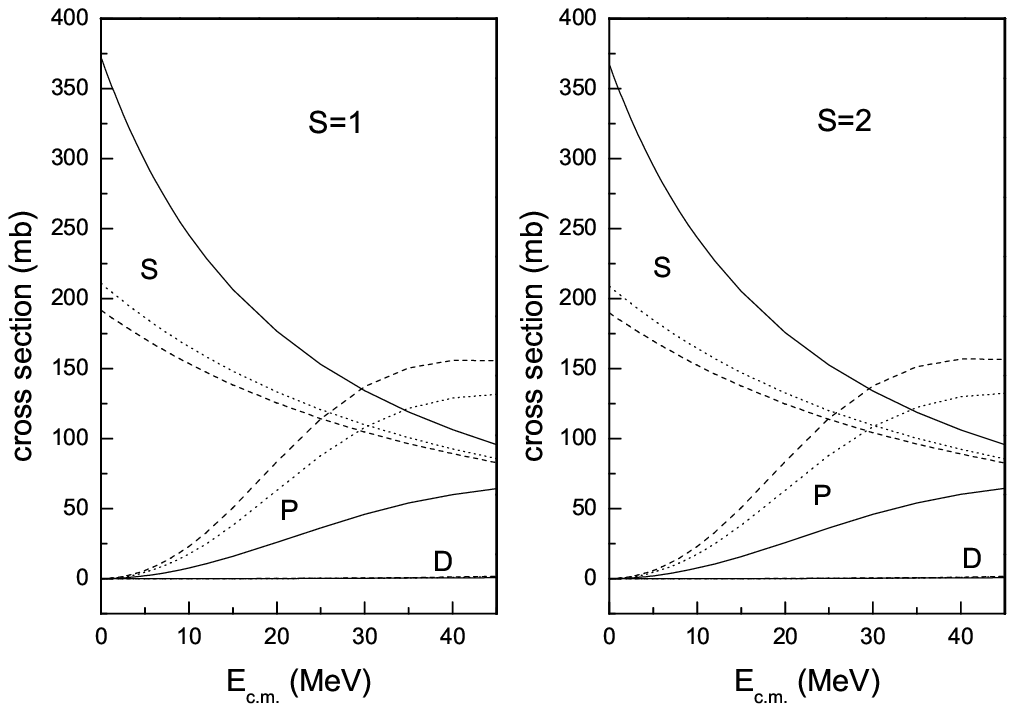,width=15.0cm,height=8.0cm} \vglue -0.5cm \caption{\small \label{pcross} The
contributions of $S$, $P$ and $D$ partial waves to $N \bar{\Omega}$ total cross sections as a function of
the energy of the center of mass motion. Same notation as in Fig. \ref{poten}.}
\end{figure}

\begin{figure}[htb]
\epsfig{file=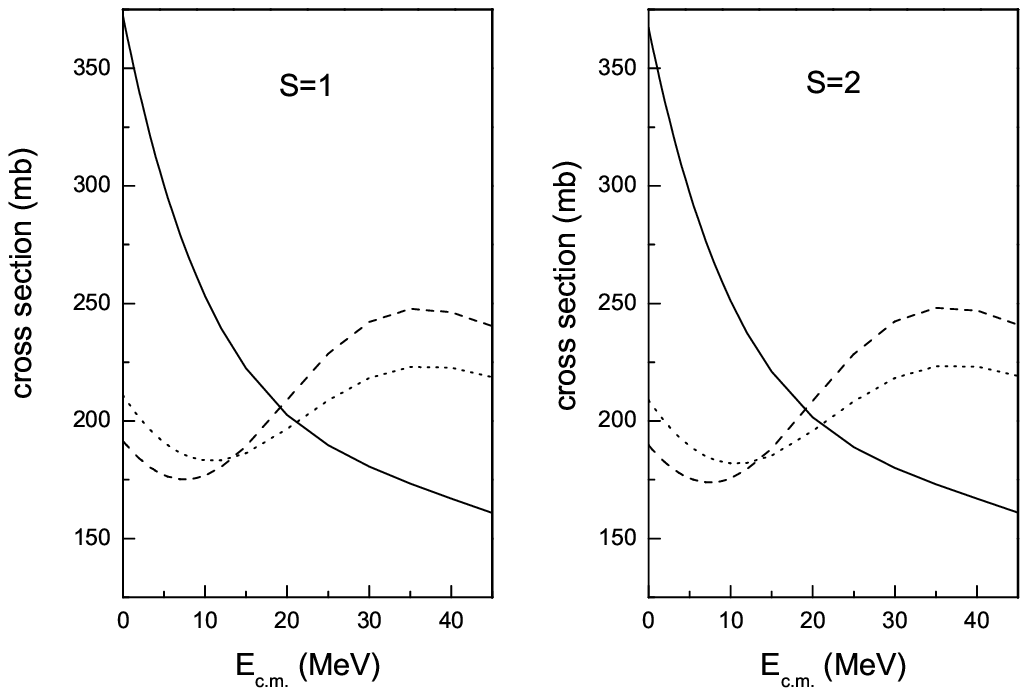,width=15.0cm,height=8.0cm} \vglue -0.5cm \caption{\small \label{cross} $N
\bar{\Omega}$ total cross sections as a function of the energy of the center of mass motion. Same
notation as in Fig. \ref{poten}.}
\end{figure}

Furthermore, the cross sections of the $N \bar{\Omega}$ elastic scattering are studied as well. The
contributions of different partial waves and the total cross sections are shown in Fig. \ref{pcross} and
Fig. \ref{cross}, respectively. From these figures, one can see that there are some differences between
the results in the chiral SU(3) quark model and those in the extended chiral SU(3) quark model. In the
very low energy region $S$ partial waves are dominantly important, and the contribution in set I is the
largest. However, with the energy enhancing, $P$-wave cross sections increase and those in sets II and
III are larger than that in set I, even larger than those of $S$ partial waves at higher energy region.
And there are nearly no contributions of $D$ partial waves. Thereby trends of curves in the extended
chiral SU(3) quark model are different from those in the chiral SU(3) quark model. The greatest
difference of the cross sections between set I and set II is about 170 mb in the very low energy region.
However in the medium energy region ($E_{cm}\approx10-30$ MeV), the cross sections are around 200-250 mb
for set I, set II and set III. It is expected that the experimental data about $N\bar{\Omega}$ elastic
scattering processes in the future will check our two chiral quark models.

The figures of scattering processes given above are the results
without considering the annihilation effect. When the effect of the
annihilation interaction is included, all the amplitudes in both
phase shifts and the total cross sections are a little higher but
the tendencies of all curves remain invariant. In addition, for
$(N\bar{\Omega})_{S=1}$ and $(N\bar{\Omega})_{S=2}$, the results of
phase shifts and cross sections are very similar. Because in our
calculation, the spin-orbit and tensor forces are neglected. When we
only consider the central force, the $\sigma$ meson exchange still
plays the dominant role, and we know it is spin independent.
Moreover, $\eta$ meson exchange includes tensor force, but it
contributes little in the $N\bar{\Omega}$ system. The spin-orbit
interaction exists in the $\sigma$ exchange, and whether it can
affect the properties of $P$ or $D$ wave deserves further study.

\section{Summary}

In summary, we perform a dynamical study of $N \bar{\Omega}$
states with spin $S=1$ and $S=2$ in the framework of the chiral
SU(3) quark model and the extended chiral SU(3) quark model by
solving the RGM equation. All the model parameters are taken to be
the values we used before, which can reasonably explain the
energies of the baryon ground states, the binding energy of
deuteron, the $NN$ scattering phase shifts, and the $YN$ cross
sections \cite{zyzhang97, lrdai03}. The numerical results show
that the $N\bar{\Omega}$ systems with both $S=1$ and $S=2$ are
bound states in these two chiral quark models. When the
annihilation effect is considered, the $N\bar{\Omega}$ system will
become more bound. This means that the annihilation effect plays
an un-negligible role in the $N \bar{\Omega}$ systems. At the same
time, the $N\bar{\Omega}$ elastic scattering phase shifts, as well
as the total cross sections are also investigated. The calculated
phase shifts are qualitatively similar in the chiral SU(3) and
extended chiral SU(3) quark models. For either the bound-state
problem or the elastic scattering processes, the results of
$(N\bar{\Omega})_{S=1}$ and $(N\bar{\Omega})_{S=2}$ are quit
alike.

It is worthy of notice that the properties of the $\bar{N}\Omega$
system are the same as those of the $N\bar{\Omega}$ one. Although
there are some problems to be further studied, such as detailed
annihilation mechanism (e.g. annihilation width or branching ratios,
etc.), the spin-orbit coupling effect for higher partial wave phase
shifts and so on, yet we still would like to emphasize that if the
qualitative feature of the $\bar{N}\Omega$ system we obtained is
right and its annihilation width is not very large, the
$N\bar{\Omega}$ system should be a very interesting one. It can be
easier to be searched in the heavy ion collision experiments,
because the abundance of $N$ is large and that of $\bar{\Omega}$,
same as $\Omega$, is not very small. More accurate study of the
structure and the properties of $N\bar{\Omega}$ system is worth
doing in the future work.

\begin{acknowledgements}
This work was supported in part by the National Natural Science
Foundation of China, Grant No. 10475087.
\end{acknowledgements}

\end{document}